\documentclass[10pt,a4paper]{article}
\usepackage[utf8]{inputenc}
\usepackage{amsmath}
\usepackage{amsfonts}
\usepackage{amssymb}
\usepackage{graphicx}
\usepackage{graphics}
\usepackage{epsfig}
\usepackage{url}
\usepackage{hyperref}
\usepackage{caption}
\usepackage{subcaption}
\usepackage{subcaption}
\usepackage{amsmath,amssymb,graphicx}

\begin{document}
\author{Gautam Kumar Saharia, Sagardeep Talukdar, 	Riki Dutta and \\ Sudipta Nandy 
\textsuperscript{a}\thanks{email: sudiptanandy@cottonuniversity.ac.in}
}

\title{Data driven localized wave solution of the Fokas-Lenells equation using modified PINN}
	
\maketitle	
	
\begin{center}
Abstract
\end{center}

We investigate data driven localized wave solutions of the Fokas-Lenells equation by using physics informed neural network(PINN). We improve basic PINN by incorporating control parameters into the residual loss function. We also add conserve quantity as another loss term to modify the PINN. Using modified PINN we obtain the data driven bright soliton and dark soliton solutions of Fokas-Lenells equation. Conserved quantities informed loss function achieve more accuracy in terms of relative $L2$ error between predicted and exact soliton solutions. We hope that the present investigation would be useful to study the applications of deep learning in nonlinear optics and other branches of nonlinear physics. Source codes are available at \url{https://github.com/gautamksaharia/Fokas-Lenells}



\vspace{5mm}
\section{Introduction}

Fokas-Lenells equation(FLE) \cite{fokas1995class,lenells2009exactly} is one of the four integrable equations which describes stable propagation of ultrashort pulses in nonlinear mediums. Significance of this equation lies on the presence of spatio-temporal term in addition to group velocity dispersion term. This equation is used to model the propagation of ultra-fast light pulses in nonlinear mediums such as pulse propagation in monomode optical fiber. The other three integrable equations are nonlinear Schr\"odinger equation (NLSE) \cite{hasegawa1973transmission, malomed2005bound}, derivative NLSE (DNLSE) \cite{kaup1978exact}, higher order NLSE (HNLSE) \cite{hirota1976n, kartashov2011solitons}. All four equations play a significant role in the study of localized waves in nonlinear optical mediums \cite{agrawal2000nonlinear}.

Nonlinear equations do not always yield stable solutions or even a solution. There are only a few analytical techniques available to deal with such solutions. While the analytical methods, namely inverse scattering transformation \cite{novikov1984theory}, direct  bi-linear method \cite{hirota1971exact, chakraborty2015bilinearization} are elegant but problem specific and are applicable to mostly integrable cases. Due to the unavailability of analytical solution in many cases, numerical methods and other approximation methods\cite{nandy2015adomian, lakshminarayanan20126} are used as an alternative technique \cite{faires2003numerical}.

However, the accuracy of the solution depends on a number of parameters, namely the number of iterations, calculation of higher order differential terms, step size etc. Moreover, some complex numerical methods are found to be very time consuming and computationally expensive. Such computationally intensive resources are not available to everyone. There is an urgent need for more general approaches to overcome limitations of analytical and numerical methods.

Deep neural network (DNN)  is one of the important discoveries in the 20th century. After a few initial successes it failed to generate much  interest but it returned in the 21st century and showed promising results when applied to a variety of fields across arts and science, namely in language processing \cite{vaswani2017attention}, image recognition \cite{gu2018recent} and many others \cite{goodfellow2016deep}. DNN generated results have  shown enough potential to be a  good alternative to the analytical and numerical methods specially in solving complex nonlinear differential equations. This is because of the most recent improvements in the computation powers and availability of abundance data. The neural network also has an obvious advantage over the analytical and numerical method in that the neural network avoids complex calculations and formulas used in the conventional  methods.

Recently Raissi et al. introduced PINN, which is computationally more efficient than traditional numerical methods, especially when the physical laws are highly nonlinear or when the geometrical domain is large and complex. After the discovery of PINN, many modifications and improvements of this method have been developed and applied to numerous fields \cite{karniadakis2021physics, jagtap2021extended, kharazmi2021hp,  samaniego2020energy, wang2021learning, jagtap2020adaptive}. While PINNs are becoming more popular, but up to this point PINNs were not  capable of accurately simulating many nonlinear dynamical systems. Some researchers addressed this drawback by modifying the loss function to increase accuracy of PINN. This resulted in overcoming some of the earlier problems seen with basic PINN. To mention a few notable approaches are, using a self adaptive loss function through the adaptive weights for each loss term \cite{xiang2022self}, using a soft attention mechanism where the adaptation weights are fully trainable and applied to each training point individually \cite{mcclenny2020self}, least squares weighted residual (LSWR) method \cite{bai2023physics}, re-formulation of PINN loss functions that can explicitly account for physical laws during model training \cite{wang2022respecting}, gradient optimization algorithm  which balances the interaction between different terms in the loss function during model training by means of gradient statistics \cite{li2022gradient}.

PINN is applied to many nonlinear equations in optics with high degree of success, namely Raissi et. al. predicted optical soliton solution of NLSE\cite{raissi2019physics}, fang et al. \cite{fang2021data} predicted femtosecond optical soliton of the high-order NLSE, peng et al. obtained rogue periodic wave of Chen–Lee–Liu equation\cite{peng2022pinn}. Data-driven solutions and parameter discovery of other nonlinear systems such as defocusing NLSE with the time dependent potential \cite{wang2021datarogue}, Manakov system \cite{pu2022data},  generalized Gross–Pitaevskii (GP) equation with PT symmetric potentials \cite{zhong2022data}, the Sasa–Satsuma \cite{luo2022data}, Yajima–Oikawa (YO) system \cite{pu2023data}, dark soliton of multi-component Manakov model \cite{jaganathan2023data}, LakshmananPorsezian-Daniel \cite{zhang2022nonlinear} have been reported.\\
\\

In all the above mentioned problems only dynamical equations along with initial-boundary conditions are used as physical information. However, the benefits of using the conserved quantity of the integrable system to the PINN method have not received enough attention. Conserved quantities are crucial in studying various optical dynamics, namely the stability of solitons,  soliton collisions and testing the stability of numerical methods. Therefore, incorporating the information of conserved quantities should improve the performance of neural networks by improving the convergence as well as generalization of neural networks. Wu et. al. \cite{wu2022prediction} has used this concept to predict optical solitons of NLSE and other notable contributions to mention are \cite{fang2022predicting, zhang2021wave}. To our knowledge, the data-driven solution of FLE  using conservation law in the PINN is not reported earlier. During our investigation we noticed that the basic PINN algorithm converges to the minimum but cannot learn the complex dynamics of FLE efficiently. Therefore, it becomes important to study complex solutions of FLE by modifying existing PINN. In this paper, we improve the PINN by modifying the loss function by incorporating a few of the conserved quantities along with other physical information into the loss function. Here, we aim to generate data driven bright soliton and dark soliton solution of FLE using modified PINN. We have shown conserved quantities informed loss function achieve more accuracy in terms of relative $L2$ error between predicted and exact solution.

This paper is organized as follows. We present a review of FLE along with its analytical bright and dark soliton solutions in section 2. In section 3 we describe the basic PINN structure and present the modified PINN structure by adding conserved quantities in the loss function. We show our results in section 4 and section 5 concludes the paper.

\section{FLE and bright dark soliton solutions}

We consider the Fokas-Lenells equation 

\begin{equation}
u_{xt}=u - 2 i \sigma |u|^2 u_x
\label{fle}
\end{equation}

where $u=u(x,t)$ is a complex valued function, where subscripts x and t denote partial differentiation with respect to $x$ and $t$. Here, $|u|^2 u_x$ accounts for the nonlinearity  and $u_{xt}$ is the spatio-temporal dispersion term.\\
\\
Under the vanishing background condition $|u| \rightarrow $  0 as $x \rightarrow \pm \infty$ we derive a bright soliton solution using Hirota bilinear method \cite{talukdar2023bilinearization}.

\begin{equation}
	u=\frac{ g_1}{f_0 + f_2}
\label{flebright1}
\end{equation}

where $$g_1= \alpha e^{\theta(x,t)}$$
$$f_0=\beta$$
$$f_2 =\beta_2 e^{\theta(x,t) + \theta^{*}(x,t)  } $$

$$\theta=p x + \frac{t}{p}+ \theta_0$$

$$\beta_2= i \frac{|\alpha|^2 |p|^2 p}{\beta^* (p+p^*)}$$

$\alpha$, $\beta$, $p$ and $\theta_0$ are arbitrary complex constants, $\alpha$ represent the polarization state, $\beta$ represents the initial central position, $p$ corresponds to the spectral parameter obtained in the Inverse scattering method and $\theta_0$ represents the initial phase. For $p = a + i b$ where $a$ and $b$ are real constant, we get the following form of bright 1-SS which we used in this paper.

\begin{equation}
u = -\frac{2 a}{a + i b} \frac{e^{\theta + i \chi}}{e^{2 \theta }- b -ia}
\label{flebright}
\end{equation}

where
$$\theta = a (x + vt)$$
$$\chi = b(x - vt)$$
$$v=\frac{1}{a^2 + b^2}$$

Again, considering FLE with non-vanishing boundary, we obtain  following \cite{matsuno2012direct}, the dark 1-SS on the background of a plane wave, that is, when
$$u \rightarrow \rho e^{ i (\kappa x - \omega t  )}, \quad x \rightarrow \pm \infty$$

The dark 1-SS is

\begin{equation}
u = \rho e^{i(\kappa x - \omega t)} \frac{1- \frac{k+b+ia}{2a} \frac{a+ i b}{a-ib} e^{2 \theta}} {1 + \frac{k + b - ia}{2a}e^{2 \theta}}
\label{fledark}
\end{equation}

where the amplitude $\rho$, frequency $\omega$ and wave number $\kappa$ of the plane wave are real constants and are constrained with the relation, namely $\omega = \frac{1}{\kappa} + 2\rho^2$. $a$ and $b$ are real constant given by  $a=\sqrt{k^3 \rho^2(1+k \rho^2)} sin(\phi)$ and $b=k \rho^2 + \sqrt{k^3 \rho^2(1+k \rho^2)} cos(\phi)$ and $0< \phi < 2\pi$

\section{PINN Deep learning method}

\subsection{PINN}

In order to write PINN for FLE, we first simplify the complex structure of FLE. We  convert $u(x,t)$ into real and imaginary parts, namely $u(x,t)$=$r(x,t)$  $+$ $i$ $m(x,t)$, where $r(x,t)$ and $m(x,t)$ are real valued functions. Writing PINN for real and imaginary part of FLE as $f_r$ and $f_m$, we have: 

\begin{equation}
f_r:= \hat r_{xt} - \hat r  -2 (\hat r^2 + \hat m^2) \hat m_x
\label{realpinn}
\end{equation}

\begin{equation}
f_m:= \hat m_{xt} - \hat m  +2 (\hat r^2 + \hat m^2)\hat r_x
\label{imagpinn}
\end{equation}

where $\hat r (x,t;w,b)$ and $\hat m (x,t;w,b)$ are the latent solutions of Neural Netowrks(NN) with the weight $W$ and bias parameters $b$, which have to be optimized to learn the exact solution $u(x,t)$.

\begin{figure}
	\centering
	\includegraphics[width=0.9\linewidth]{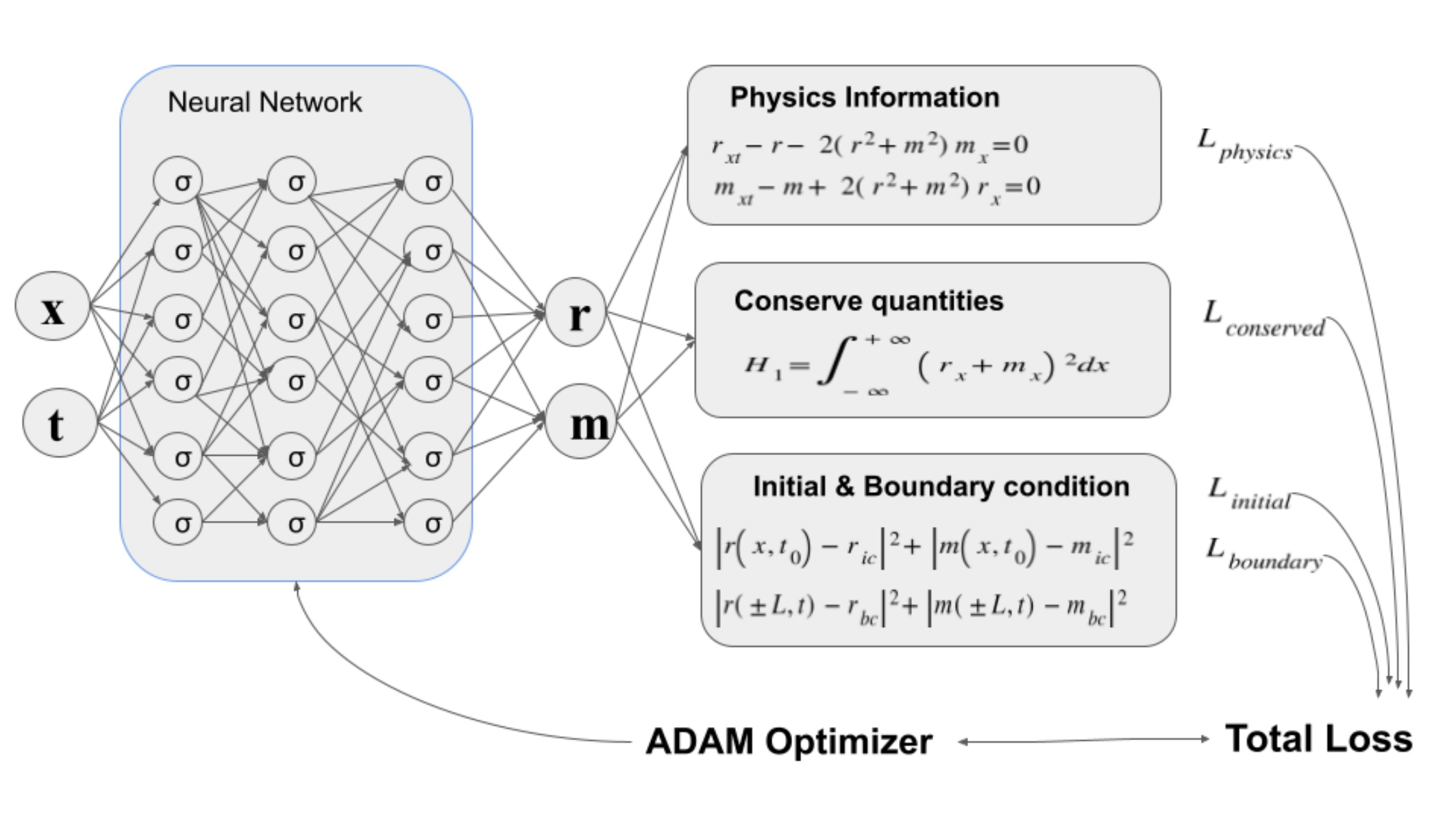}
	\caption{Schematic diagram of PINN. A NN consists of an input and an output layer and some number of hidden layers. $\sigma$ represents activation functions in all layer. The input of the NN goes through all the hidden layers. Outputs of the NN, namely $r$ and $m$ are considered inputs for the initial and boundary condition loss functions, the physics equation loss function, and the conserve quantity loss function. The total loss function is the combination of all these loss functions, which are minimized by the ADAM optimizer.}
	\label{fig:untitled-presentation}
\end{figure}

We construct a feed-forward  NN, which consists of an input layer, M number of hidden layers and an output layer as shown in Figure \ref{fig:untitled-presentation}. The input layer takes the coordinates $(x,t)$ as input, multiplies them with weight $W$ and adds bias $b$ to it. Before sending them as an input to the next layer, we apply an activation function $\sigma$, namely tanh to add non-linearity in the output. The network is called feed-forward because each hidden layer of the NN receives input from the previous layer. We use Glorot Normal initialization to randomly initialize the network weight $W$ and bias term $b$. The final NN representation is given by,

\begin{equation}
u(X, \Theta) = \sigma(\sigma (\sigma(W_0 X + b_0)W_1 + b_1)W_2 + b_2)......
\label{nn1}
\end{equation}

$u(X, \Theta) = [\hat r (x,t;w,b), \hat m (x,t;w,b)]$ is the output of the NN, and $X=(x, t)$ is the input to the NN,  $\Theta$ = $\left\lbrace W^k, b^k\right\rbrace^M_{k=1} $ represent trainable parameters in the NN and $\sigma$ represents activation function. Our goal is to optimize these trainable parameters so that $\hat u(x,t)$ satisfy the FLE and PINN $f_r$ and $f_m$ become minimum, such that the output of the NN approximates the solution of FLE, i.e. $\hat u(x,t;w,b) \approx u(x,t)$.

PINN $f_r$ and $f_m$ share same parameters with the NN $ u(X, \Theta)$ therefore, these common parameters are trained by minimizing the following loss function of the network:

\begin{equation}
Loss(\Theta) = Loss_{r_0} + Loss_{m_0} + Loss_{r_b} + Loss_{m_b}+ Loss_{f_r} + Loss_{f_m}
\label{totalfleloss}
\end{equation}

where, $Loss_{r_0}$, $Loss_{m_0}$, $Loss_{r_b}$, $Loss_{m_b}$, $Loss_{f_r}$ and $Loss_{f_m}$ are defined by

\begin{equation}
Loss_{r_0}= \frac{1}{N_0} \sum_{i=1}^{N_0}|\hat r(x^i_0, t^i_0)-r^i_0|^2 , \quad Loss_{m_0}= \frac{1}{N_0} \sum_{i=1}^{N_0}|\hat m(x^i_0, t^i_0)-m^i_0|^2
\label{fleinitailloss}
\end{equation}

\begin{equation}
Loss_{r_b}= \frac{1}{N_b} \sum_{i=1}^{N_b}|\hat r(x^i_b, t^i_b)-r^i_b|^2, \quad Loss_{m_b}= \frac{1}{N_b} \sum_{i=1}^{N_b}|\hat m(x^i_b, t^i_b)-m^i_b|^2
\label{fleboundaryloss}
\end{equation}

\begin{equation}
Loss_{f_r}=\frac{1}{N_f} \sum_{i=1}^{N_f}|f_r(x^i_f, t^i_f)|^2, \quad Loss_{f_m}=\frac{1}{N_f} \sum_{i=1}^{N_f}|f_m(x^i_f, t^i_f)|^2
\label{flepdeloss}
\end{equation}

Here, $\left\lbrace x^i_0, t^i_0; r^i_0, m^i_0 \right\rbrace^{N_0}_{i=1} $ denotes initial data points, $\left\lbrace x^i_b, t^i_b; r^i_b, m^i_b \right\rbrace^{N_b}_{i=1} $ corresponds to collocation points on the boundary data, and $\left\lbrace x^i_f, t^i_f \right\rbrace^{N_f}_{i=1} $ represent the collocation points on $f_r$ and $f_m$. $Loss_{r_0}$ and $Loss_{m_0}$ corresponds to the loss on the initial data, $Loss_{r_b}$ and $Loss_{m_b}$ enforce the vanishing boundary conditions, and  $f_r$ and $f_m$ penalizes the FLE for not being satisfied on the collocation points. These training data were obtained from an exact solution, considering their initial and boundary conditions. Collocation points are randomly chosen from a uniform distribution in the computational domain i.e. $x \in [-L,L]$ and $t \in [t_0, t_1]$. 

With the help of the ADAM optimizer \cite{kingma2014adam}, we minimize the loss function by optimizing the trainable parameters $\Theta$ = $\left\lbrace W^k, b^k\right\rbrace^M_{k=1} $. Thus, solving the FLE is now transformed into an optimization problem. The global minimum of the loss function corresponds to the solution of the FLE subject to particular initial and boundary conditions.

\subsection{ Loss function with conserve quantity}

Although the basic PINN method converges to a minimum value, it cannot learn the exact soliton solution of the FLE. To address this problem, we modify the loss function by multiplying some real coefficient $\gamma$ in the physics informed loss function, i.e $f_r$ and $f_m$. Physics informed loss refers to imposing FLE residuals in the total loss functions to regularize NN training. The modified loss function is given as follows:

\begin{equation}
Loss(\Theta) = Loss_{r_0} + Loss_{m_0} + Loss_{r_b} + Loss_{m_b}+ \gamma(Loss_{f_r} + Loss_{f_m})
\label{wloss}
\end{equation}
where $\gamma$ determines how much of a physics informed loss function is to be contributed to the total loss function.

Embedding the knowledge of physical information, namely energy conservation law, momentum conservation law etc into a NN as a loss function improves the PINN to capture the right solution and generalize better result even with a small number of training data. Therefore, we further improve the loss function by introducing additional physical information from the FLE. Being an integrable equation, FLE has many conserved quantities. We use conserve quantities of FLE in the PINN method to improve the loss function by adding them as additional loss terms which have to be minimized during training. 

The first few conserved quantities for FLE are given below, including both positive and negative hierarchy, which are obtained by solving the methods described in \cite{talukdar2023bilinearization}. 

\begin{equation}
	H_1 = \int^{+\infty}_{-\infty} |u_x|^2 dx
	\label{h1}
\end{equation}

\begin{equation}
	H_3 = \int^{+\infty}_{-\infty} (|u_x|^4 - i u^*_x u_{xx})dx
	\label{h3}
\end{equation}

\begin{equation}
	H_5=\int^{+\infty}_{-\infty}(-u^*_x u_{xxx} + 2 |u_x|^6  -3 i |u_x|^2 u^*_x u_{xx})dx
	\label{h5}
\end{equation}

\begin{equation}
	H_{-1} = \int^{+\infty}_{-\infty}-iu^*_x u dx
	\label{h-1}
\end{equation}

\begin{equation}
	H_{-3} = \int^{+\infty}_{-\infty}-(|u|^2 + i |u|^2 u u^*_x)dx
	\label{h-3}
\end{equation}

Since the temporal derivative of conservative quantity must be minimum,  we write

\begin{equation}
\frac{\partial}{\partial t} H_i=0 , \quad i=1, 3, 5, -1, -3
\end{equation}

\begin{equation}
L_{conserve} = \sum_{i=1,3, 5, -1, -3}||\frac{\partial}{\partial t} \hat H_i||^2
\end{equation}

Where $L_{conserve}$ corresponds to loss on conserve quantities of FLE. Thus, improved loss function with conserved quantity:
\begin{equation}
Loss(\Theta) = Loss_{r_0} + Loss_{m_0} + Loss_{r_b} + Loss_{m_b}+ \gamma_1(Loss_{f_r} +
 Loss_{f_m}) + \gamma_2 L_{cnsr}
\label{closs}
\end{equation}

where $\gamma_1$ and $\gamma_2$ are two real-valued coefficients that determine the contribution from the physics loss function and conserved quantity loss function respectively. Improved loss functions present more precise soliton behaviors with fewer sample points.

\subsection{Algorithm}

\textbf{Step 1}: Specification of training data, 
initial training data: $(x^i, t^i, r^i, m^i)^{N_0}_{i=1}$; boundary training data: $(x^j, t^j, r^j, m^j)^{N_b}_{j=1}$; residual training points: $(x^l_f, t^l_f)^{N_f}_{l=1}$

\textbf{Step 2}: Construct the NN $u_{NN}(\Theta)$ with random initialization of trainable parameters $\Theta \in \left\lbrace W, b\right\rbrace $

\textbf{Step 3}: Construct the PINN  by substituting surrogate $u_{NN}$ into the governing equation.

\textbf{Step 4}: Specification of the loss function, that include the weighted loss function and conserved quantity.

\textbf{Step 5} : Minimize the loss function to find best parameters $\Theta$ using ADAM Optimizer

\section{Results}
\subsection{Bright Soliton}

We consider the initial condition $u(x,-1)=g(x)$ for $a=1$ and $b=1$

\begin{equation}
g(x)= -\frac{2 }{1 + i } \frac{ {e^{(x - \frac{1}{2}) + i (x + \frac{1}{2})}}} {e^{ 2(x - \frac{1}{2} ) }- 1 -i} \quad x \in [-5,5]
\end{equation}

and the vanishing boundary condition, namely

$$u(x=-5,t)=u(x=5, t)=0 ,\quad t \in [-1,1]$$	

Here, we chose $ x \in [-5,5]$ and $t \in [-1,1]$ as the spatial and temporal interval.

Training data $\left\lbrace x^i_0, -1 \right\rbrace^{N_0}_{i=1}  $ for initial condition consists of $N_0=100$ data points, randomly drawn from uniform distribution over the half open interval $[-5,5)$, similarly, $N_b=100$ data points $\left\lbrace \pm 5, t^i_b \right\rbrace^{N_b}_{i=1} $ are drawn from uniform distribution over the half open interval $[-1,1)$ to enforce the vanishing boundary condition. Moreover, we have selected $N_f=1000$ number of randomly sampled collocation points to enforce the eq. \ref{fle} inside the computational domain. All these randomly sampled points are drawn from the uniform distribution over the spatial and temporal intervals, $ x \in [-5,5]$ and $t \in [-1,1]$.

To obtain a data-driven bright soliton, we construct a feed-forward NN with eight hidden layers with 40 neurons in each layer and hyperbolic tan as the activation function. Minimizing the loss function by optimizing all the learnable parameters using the ADAM optimizer with a learning rate of 0.001, we approximate the bright 1-SS.

Considering $\gamma=0.001$ in eq. \ref{wloss} we obtain $L_2$ error,  where total loss= 0.00085254 and physics loss = 0.79478168 after 30,000 iterations in time 1187.6953 seconds. Again considering $\gamma_1=0.001$ and $\gamma_2=0.00001$  multiplying with $H_{-1}$ conserved loss function in eq. \ref{closs} we get $L2$ error where total loss = 0.00072432 and physics loss = 0.647380 after 30,000 iterations in 4393.4230 seconds. Thus, we find that inclusion of conserved quantities to the loss function enables us to minimization of loss more effectively than when it is excluded.

\begin{figure}[htbp]
\centering
\begin{subfigure}{.3\textwidth}
	\centering
	\includegraphics[width=0.9\linewidth]{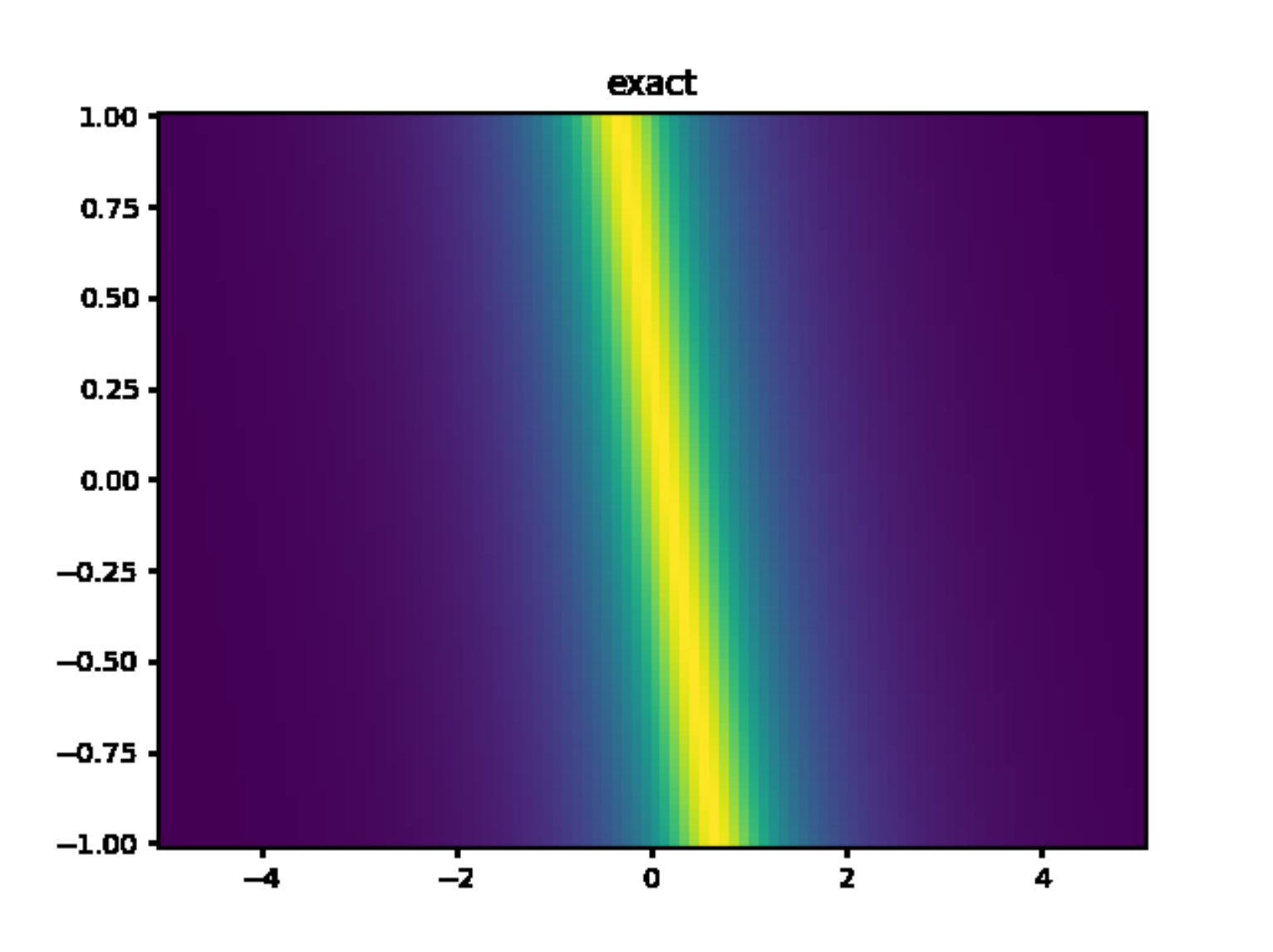}
	\caption{Exact}
	\label{fig:exactbrigh1ss}
\end{subfigure}
\hfil
\begin{subfigure}{.3\textwidth}
	\centering
	\includegraphics[width=0.9\linewidth]{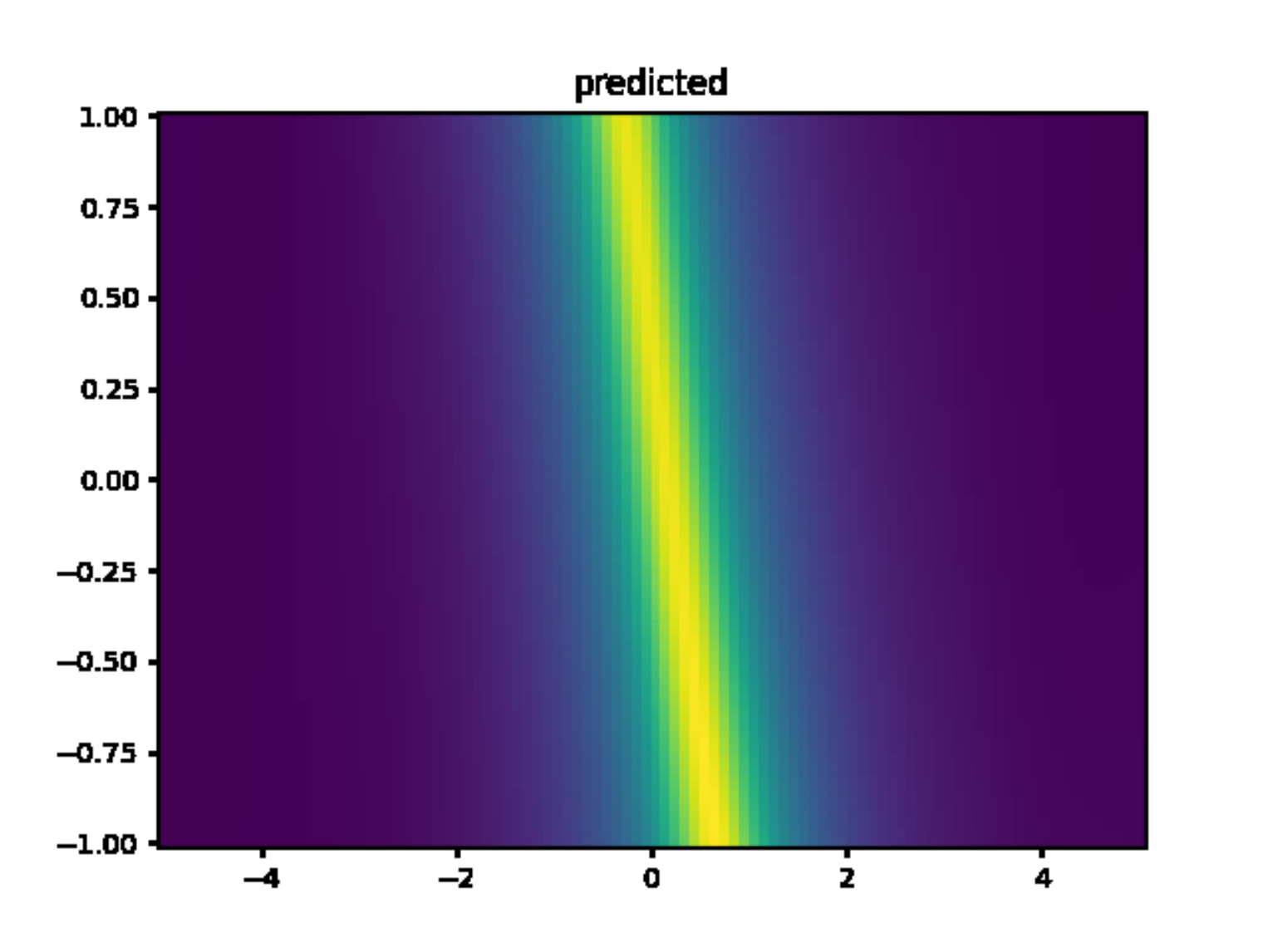}
	\caption{Data-driven}
	\label{fig:datadrivenbrigh1ss}
\end{subfigure}
\hfil
\begin{subfigure}{.3\textwidth}
	\centering
	\includegraphics[width=0.9\linewidth]{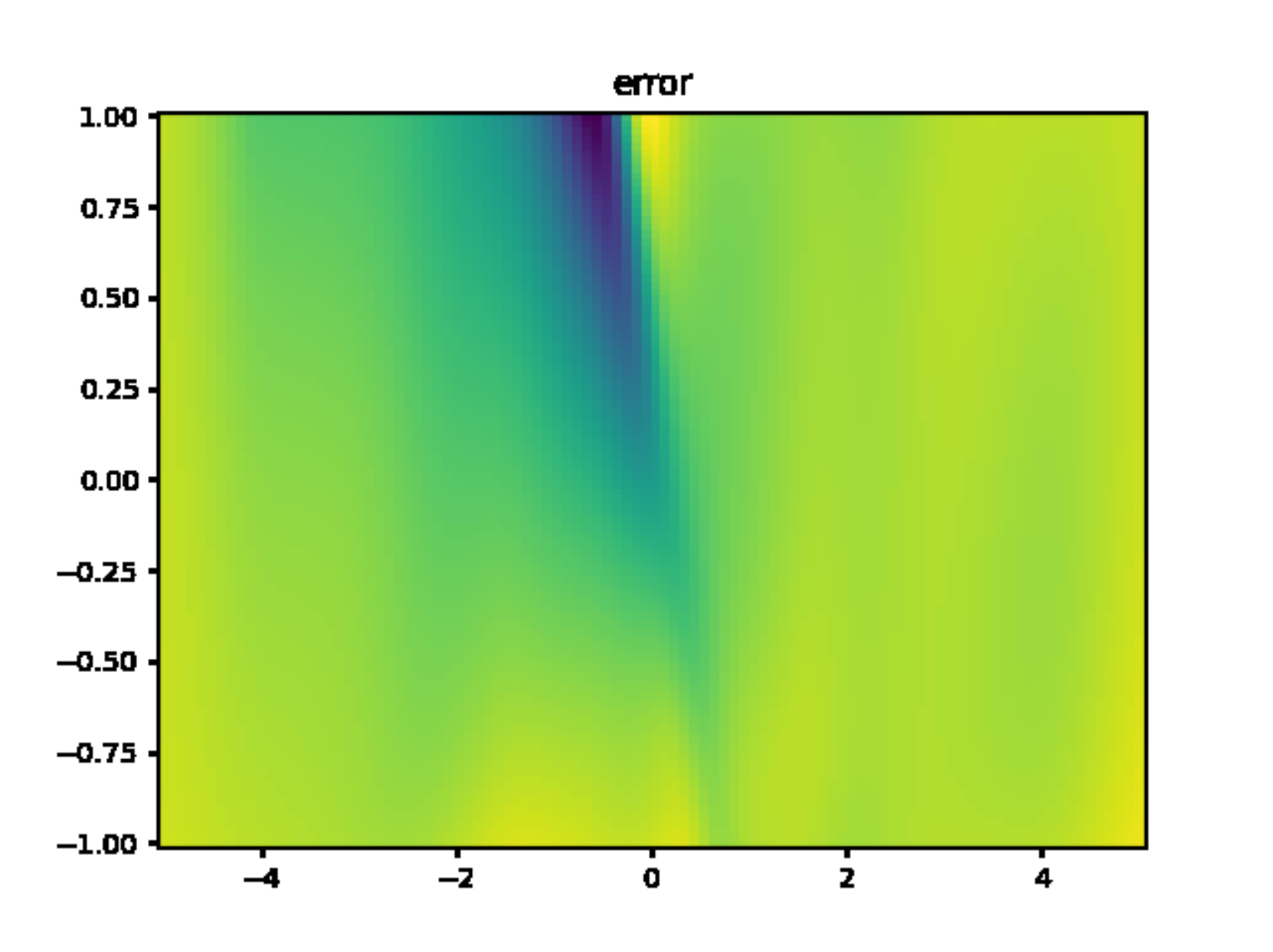}
	\caption{Error}
	\label{fig:error}
\end{subfigure}
\caption{Exact and data-driven bright 1-SS and error between them}
\label{bright1ss}
\end{figure}

\begin{figure}[htbp]
\centering
\begin{subfigure}{.3\textwidth}
	\centering
	\includegraphics[width=0.9\linewidth]{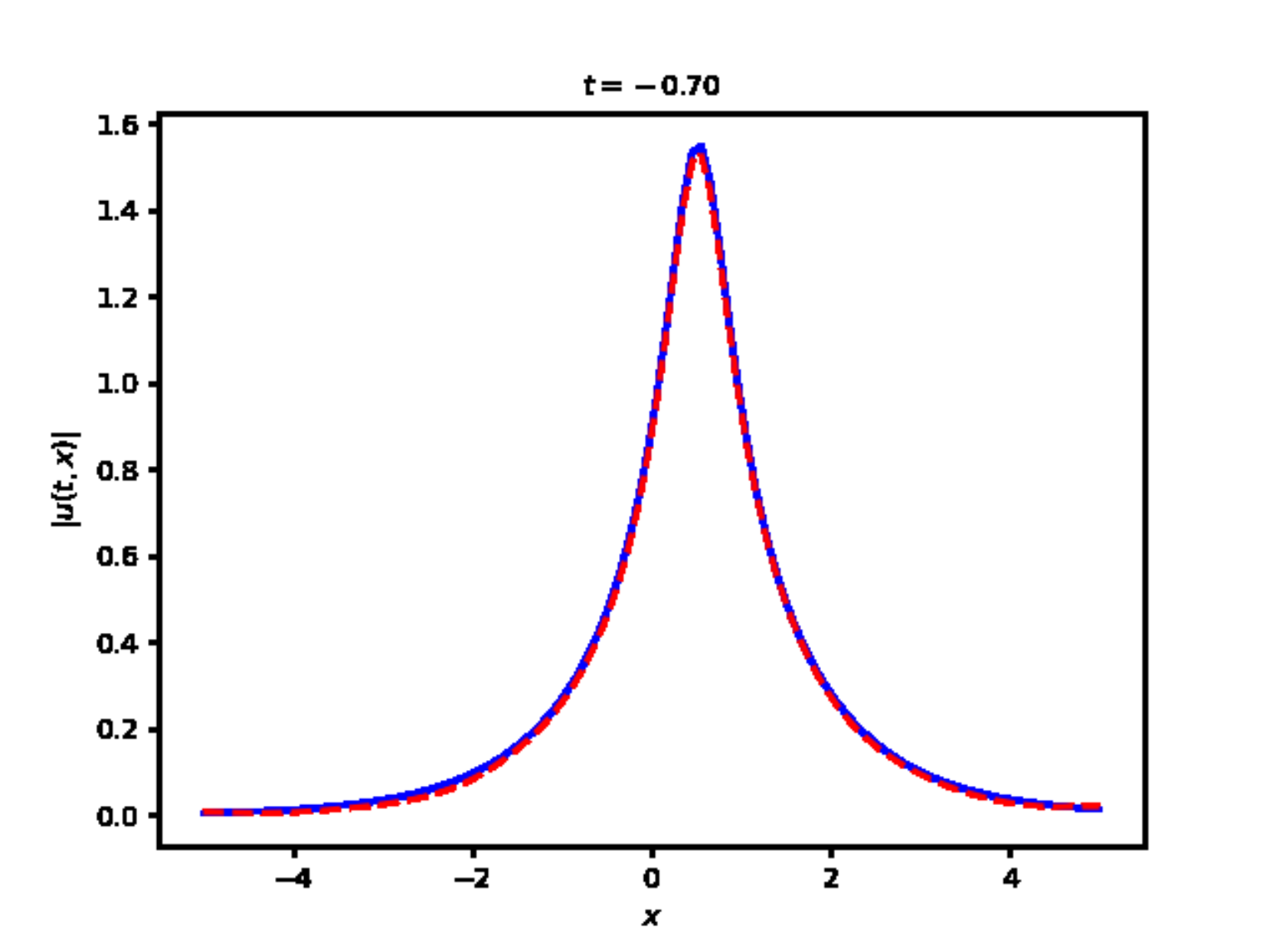}
	\caption{at t = -0.70}
	\label{fig:compare33bright1ss}
\end{subfigure}
\hfil
\begin{subfigure}{.3\textwidth}
	\centering
	\includegraphics[width=0.9\linewidth]{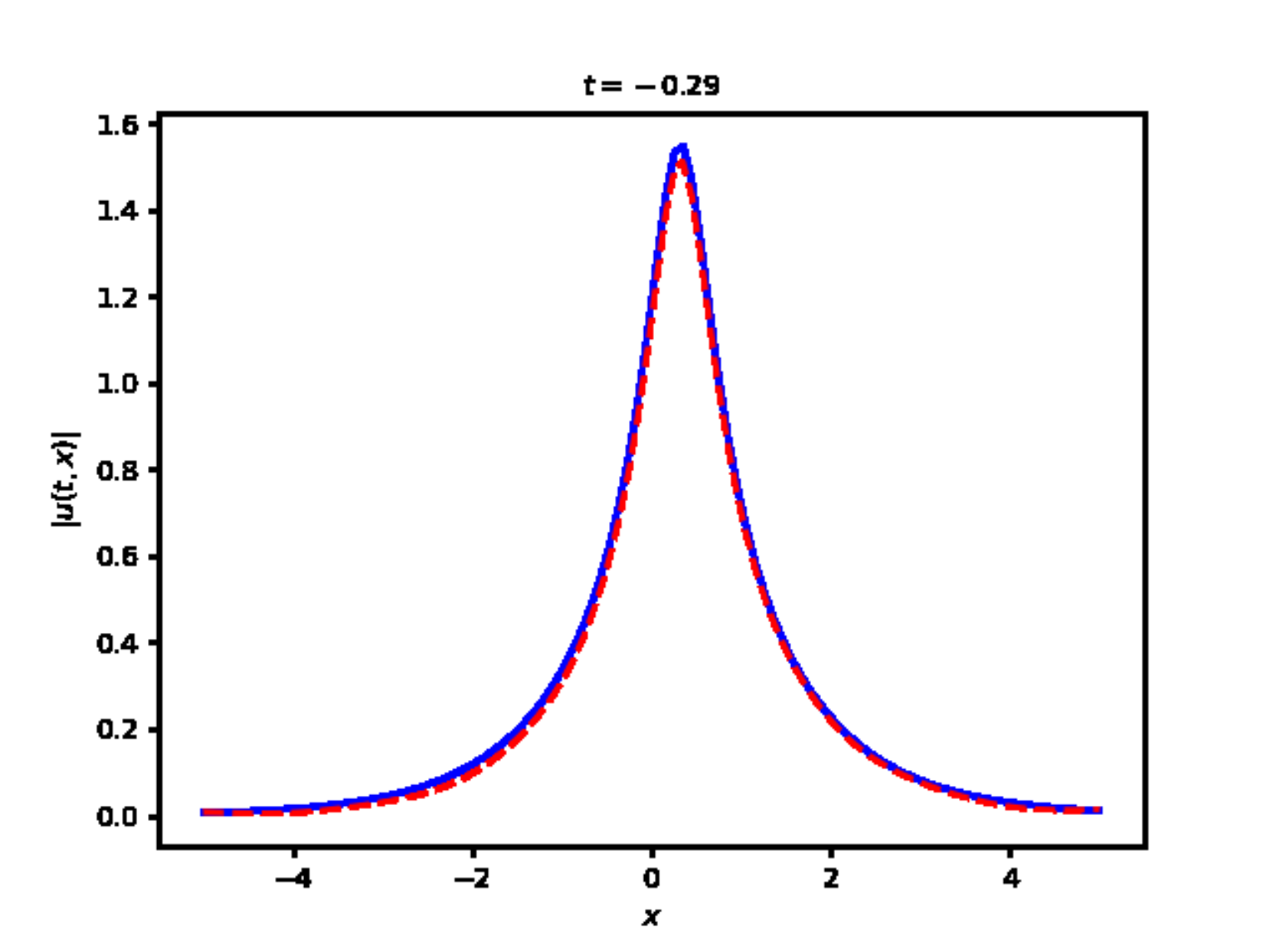}
	\caption{at t = -0.29}
	\label{fig:compare22bright1ss}
\end{subfigure}
\hfil
\begin{subfigure}{.3\textwidth}
	\centering
	\includegraphics[width=0.9\linewidth]{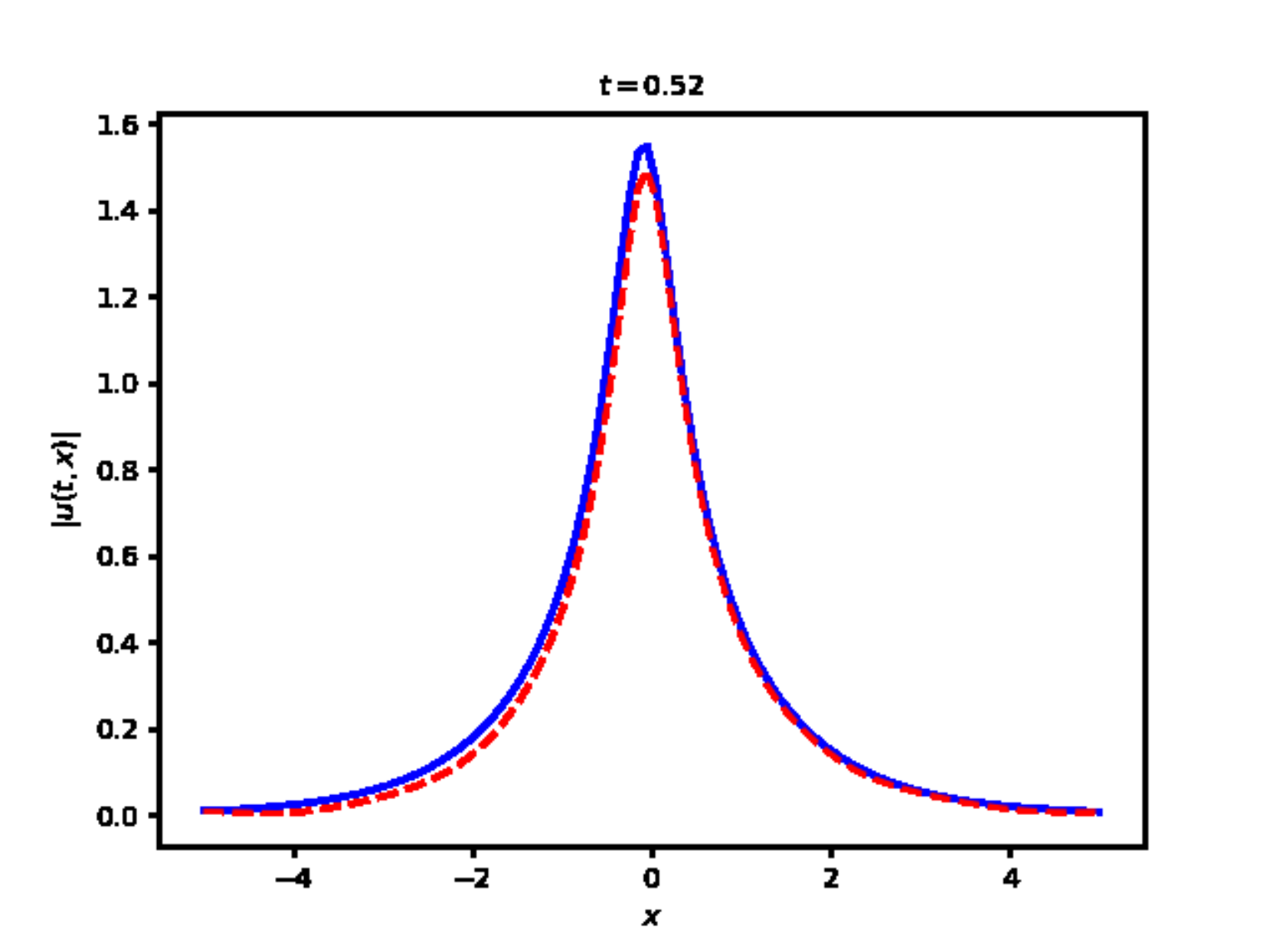}
	\caption{at t = 0.52}
	\label{fig:compare11bright1ss}
\end{subfigure}
\caption{Comparison between exact and data driven bright 1-SS at three different time instance (a)at t = -0.70, (b) at t = -0.29, (c) at t = 0.72}
\label{bright1ss_compare}
\end{figure}

The results are demonstrated in Figure \ref{bright1ss}  and Figure \ref{bright1ss_compare}. Figure. \ref{bright1ss} (a) shows density plot of exact bright 1-SS, Figure.\ref{bright1ss}(b) shows density plot of data-driven bright 1-SS and Figure \ref{bright1ss}(c) shows density plot of error between exact and data-driven bright 1-SS. Figure \ref{bright1ss_compare} shows a comparison of the soliton at different time instance  (i) at t = -0.70, (ii) at t = -0.29 (iii) at t = 0.52.

\subsection{Dark Soliton}


Initial value condition for dark soliton $u(x,-1) = g(x) $, considering $a=\sqrt{2}$, $b=1$, $\kappa=1$, $\rho=1$ and $\omega=3$.
$$g(x) = e^{i(x+3)} \frac{1- \frac{2+i}{2} \frac{1+ i }{1-i} e^{2x-1 }} {1 + \frac{2 - i}{2}e^{2 x-1}} \quad ,x \in [-5,5]$$ 

and the boundary conditions

$$u(-5, t)=u(5,t)=1 , \quad  \text{t} \in [-1,1]$$

Here , we chose $ x \in [-5,5]$ and $t \in [-1,1]$ as the spatial and temporal interval.

The training data set is obtained considering the exact solution at the initial boundary data i.e. at $x =-5$, $x=5$, and $t=-1$ dividing the spatial region [-5, 5] into $N_b=200$ data points and temporal region [-1,1] into $N_0=100$ points. We have also selected a $N_f=2000$ number of randomly sampled collocation points to enforce the eq. \ref{fledark} inside the computational domain. All these randomly sampled points are drawn from the uniform distribution over the spatial and temporal intervals $ x \in [-5,5]$ and $t \in [-1,1]$.

We use the 9 hidden layer deep PINN with 60 neurons per layer and a hyperbolic tan activation function to obtain data driven dark 1-SS. To minimizing the loss function, we use the ADAM optimizer with a learning rate 0.001. We obtain dark 1-SS with total loss= 0.00221427 and physics loss = 1.88655448 after 30,000 iterations with coefficient value of $\gamma_1 = 0.001$ to physics loss function and $\gamma_2 = 0.001$ to the $H_{-3}$ conserved quantities loss function.

\begin{figure}[htbp]
	\centering
	\begin{subfigure}{.3\textwidth}
		\centering
		\includegraphics[width=0.9\linewidth]{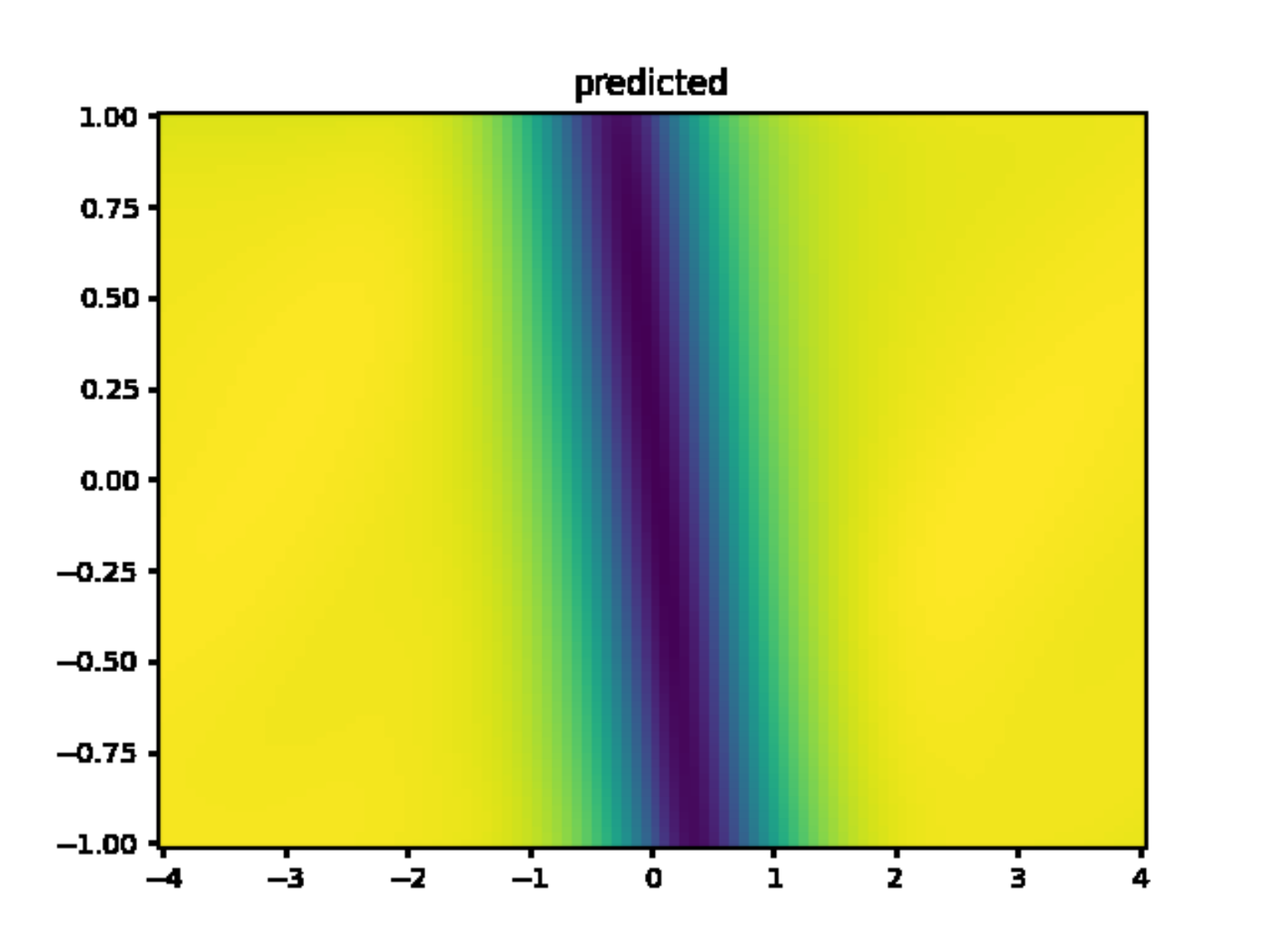}
		\caption{data driven}
		\label{fig:densitydatadrivendarkfle}
	\end{subfigure}
	\hfil
	\begin{subfigure}{.3\textwidth}
		\centering
		\includegraphics[width=0.9\linewidth]{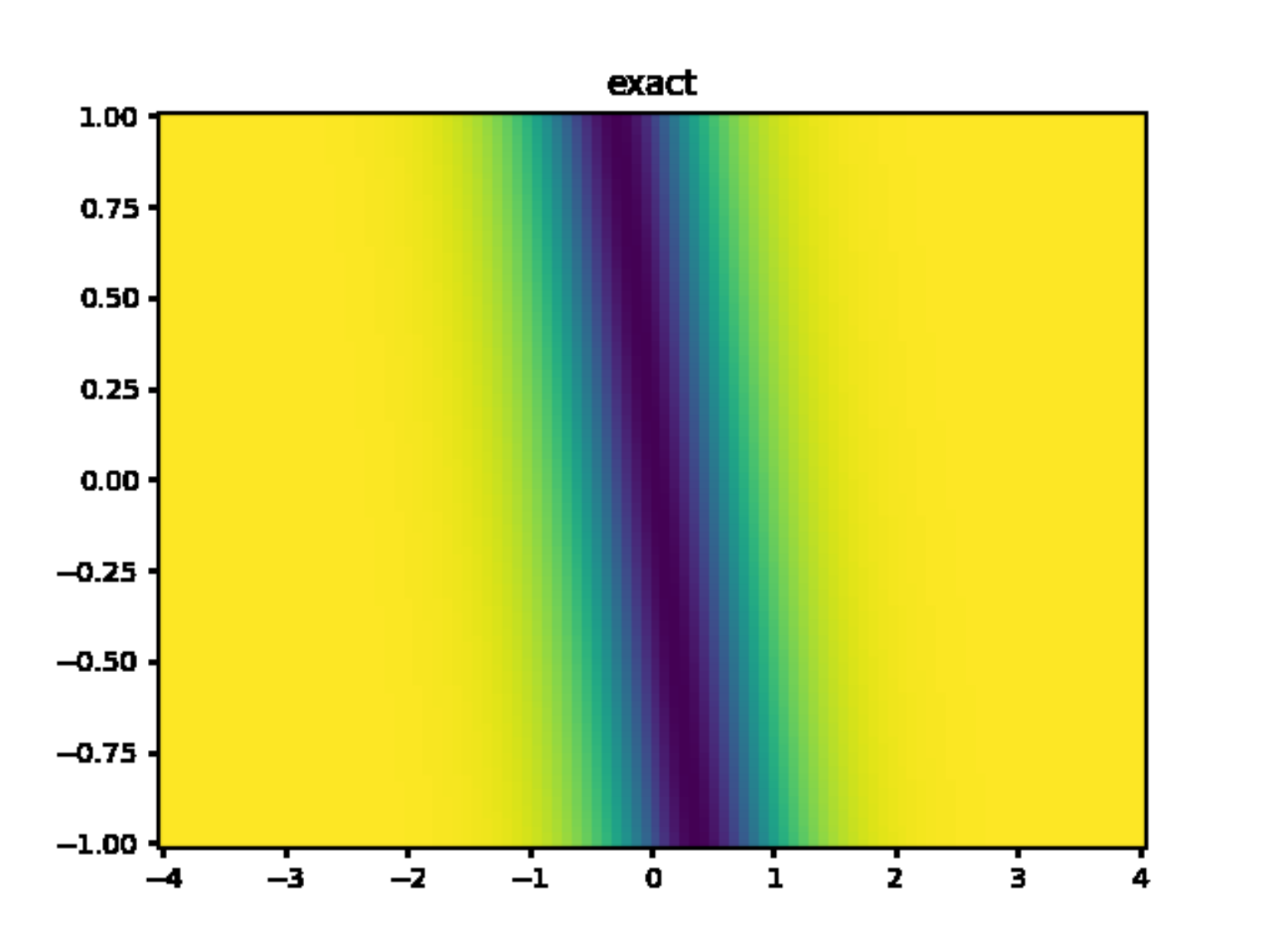}
		\caption{exact}
		\label{fig:densityexactdarkfle}
	\end{subfigure}
	\hfil
	\begin{subfigure}{.3\textwidth}
		\centering
		\includegraphics[width=0.9\linewidth]{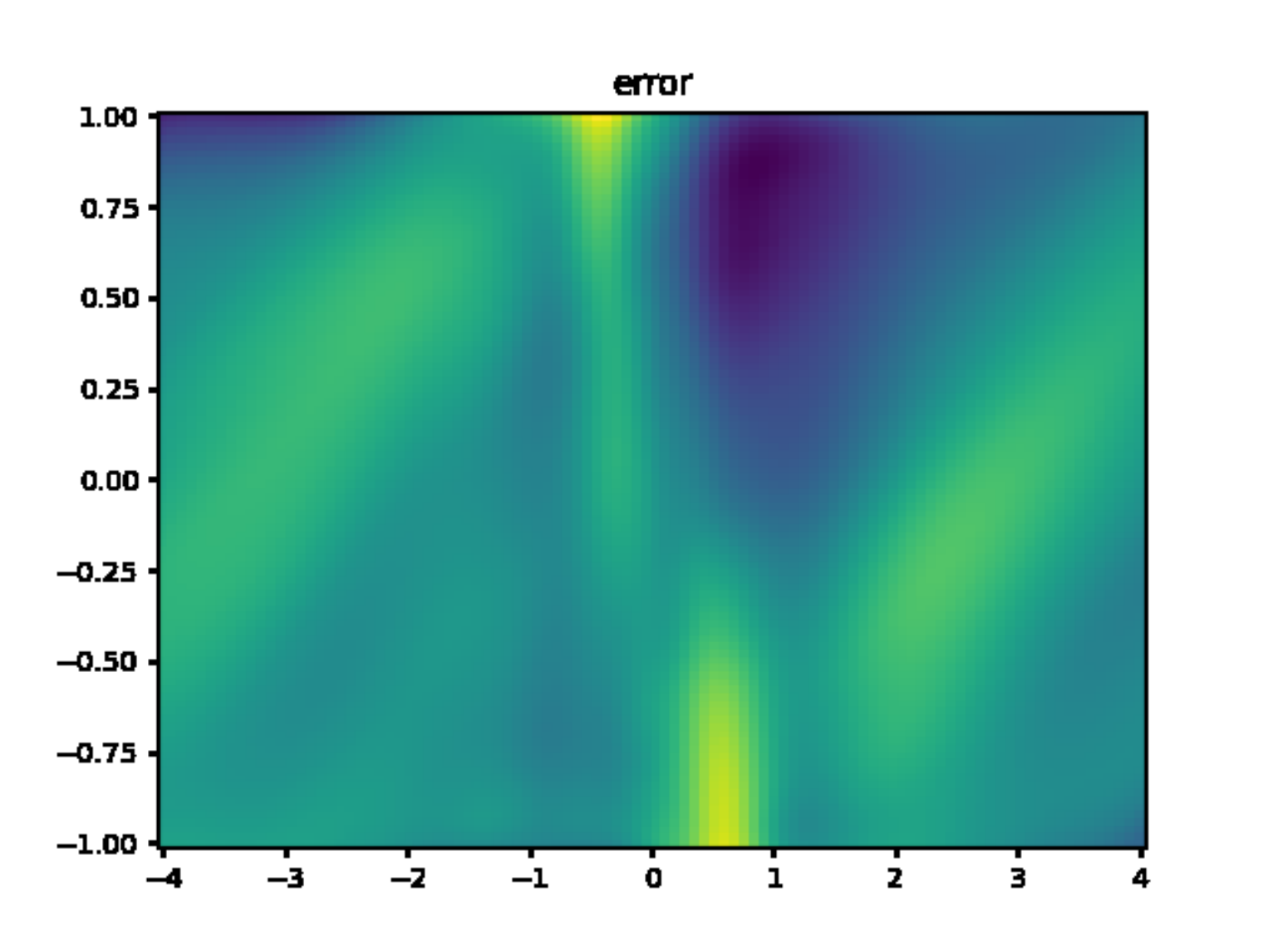}
		\caption{error}
		\label{fig:errordarkfle}
	\end{subfigure}
	\caption{Data-driven and exact dark 1-SS and error between them }
	\label{dark1ss}
\end{figure}

\begin{figure}[htbp]
	\centering
	\begin{subfigure}{.3\textwidth}
		\centering
		\includegraphics[width=0.9\linewidth]{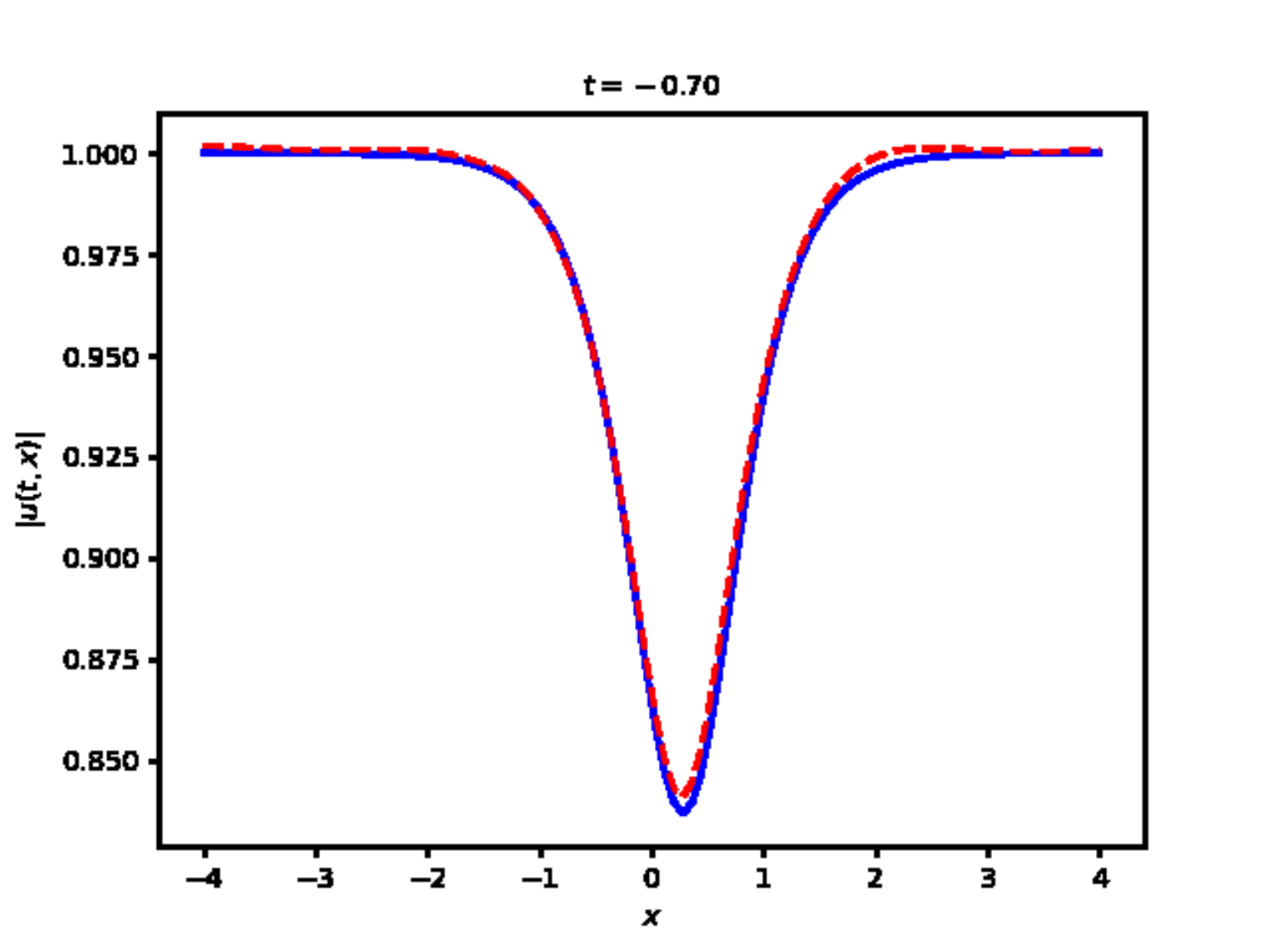}
		\caption{at t = -0.70}
		\label{fig:compare33darkconserve-1}
	\end{subfigure}
	\hfil
	\begin{subfigure}{.3\textwidth}
		\centering
		\includegraphics[width=0.9\linewidth]{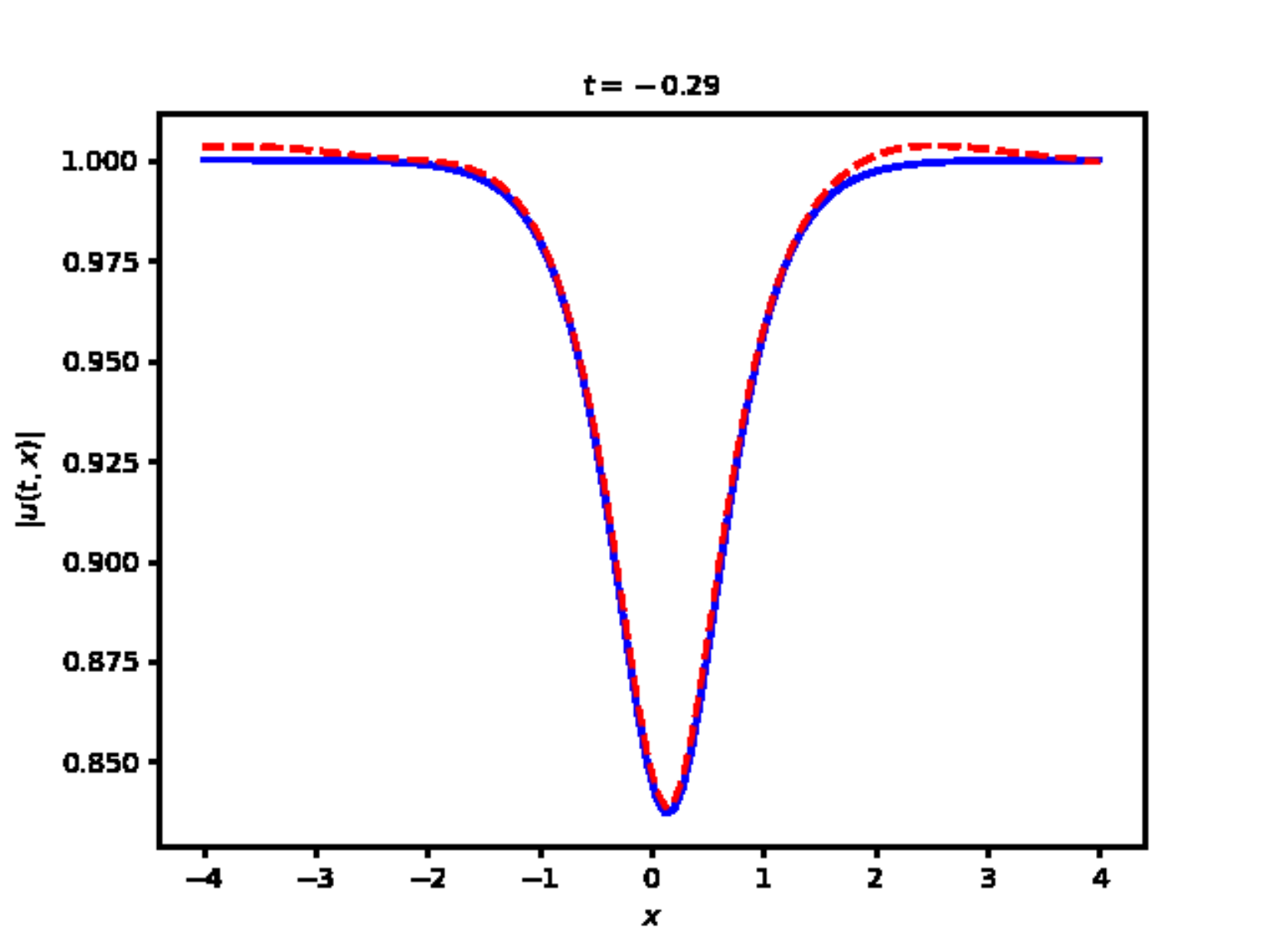}
		\caption{at t = -0.29}
		\label{fig:compare22darkconserve-1}
	\end{subfigure}
	\hfil
	\begin{subfigure}{.3\textwidth}
		\centering
		\includegraphics[width=0.9\linewidth]{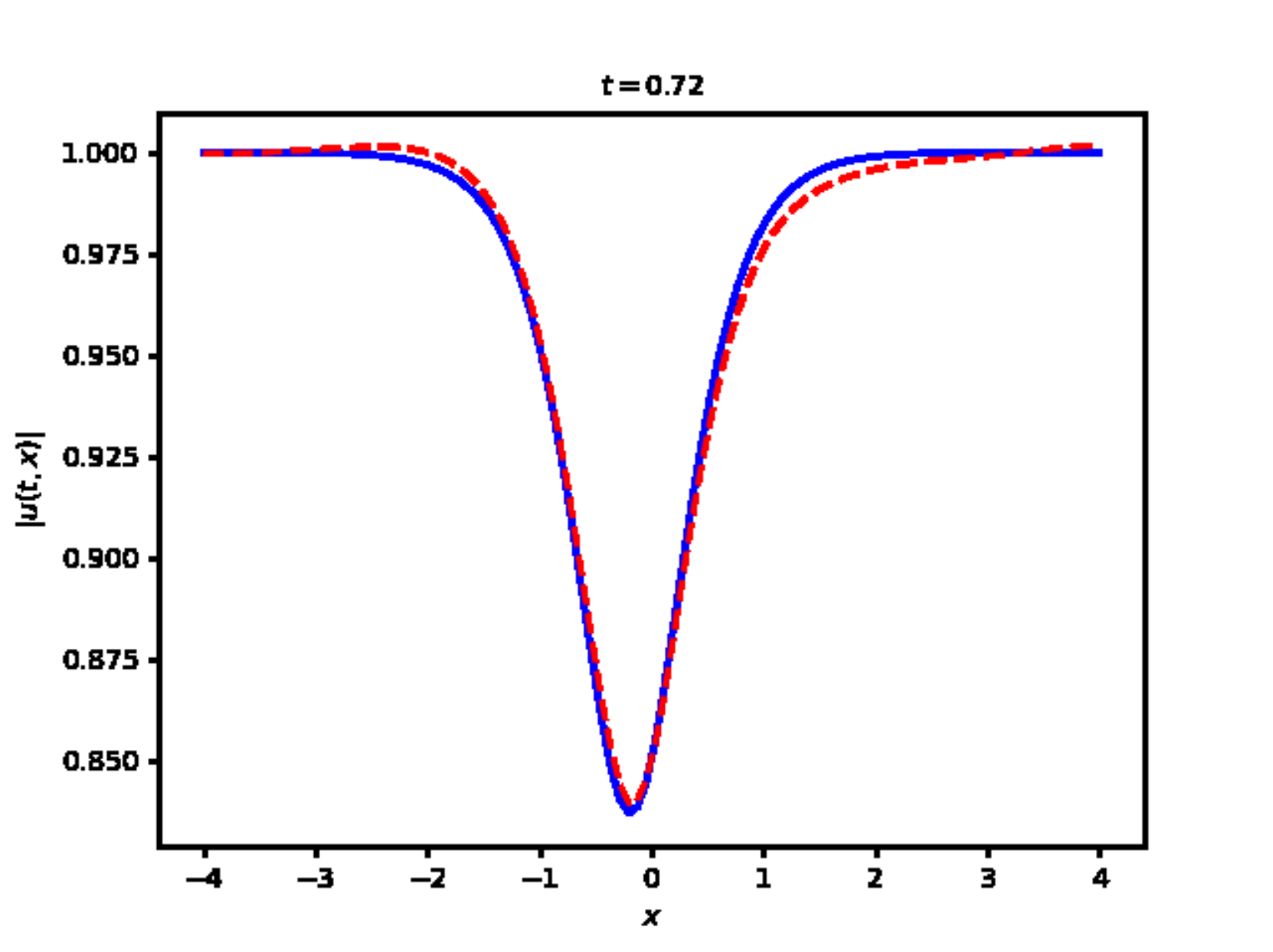}
		\caption{at t =0.72}
		\label{fig:compare11darkconserve-1}
	\end{subfigure}
	\caption{Comparison between exact and data driven dark soliton solution at three different time instance (a)at t = -0.70, (b) at t = -0.29, (c) at t = 0.72}
	\label{dark_compare}
\end{figure}

The results are demonstrated in Figure. \ref{dark1ss}  and \ref{dark_compare}. Figure. \ref{bright1ss} (a) shows density plot of exact dark 1-SS, Figure. \ref{bright1ss} (b) shows density plot of data-driven dark 1-SS and Figure. \ref{bright1ss}  (c) shows density plot of error between exact and data-driven dark 1-SS. We present a comparison between the exact and data-driven dark soliton solution at three different time instants (i) $t = -0.29$, (ii) $t=-0.70$ and (ii) $t=0.72$ in Figure \ref{dark_compare}.

\section{Conclusion}
We improve basic PINN by incorporating control parameters into the residual loss function. We add the conserved quantities of FLE into the loss function to modify the PINN. We have obtained both bright and dark 1-SS of FLE using an improved PINN. Although PINN can predict bright soliton solutions with great accuracy, for dark soliton, there is still scope for improving its accuracy. We find that using conserved quantities of FLE as another loss term helps us to obtain a data-driven soliton solution with greater accuracy. Therefore, incorporating the information of conserved quantities should improve the performance of the NN by improving its convergence as well as generalization.

\section*{Acknowledgement}
G.K. Saharia acknowledges Google Colab and Kaggle for their free GPU services.
S. Talukdar and R. Dutta acknowledges DST, Govt. of India for Inspire fellowship,
grant nos. DST Inspire Fellowship 2020/IF200278; 2020/IF200303.



\end{document}